\documentclass[useAMS,usedcolumn,usegraphicx,usenatbib]{mn2e}

\usepackage{times}
\usepackage{amsmath}                  % mathematischer Zeichensatz
\usepackage{amssymb}                  % mathematische Symbole
\usepackage{graphicx}
\usepackage{units}
\usepackage{dcolumn}
\usepackage{longtable}
\usepackage{lscape}
\usepackage{booktabs}
\usepackage{natbib}
\usepackage{aas_macros}
\usepackage[draft,pdftex,pdfpagemode={UseOutlines},bookmarks,bookmarksopen,colorlinks,linkcolor={blue},citecolor={green},urlcolor={red}]{hyperref}
\makeatletter
\newcolumntype{d}[1]{>{\DC@{,}{{.}}{#1}}c<{\DC@end}}
\newcolumntype{o}[1]{>{\DC@{+}{\pm}{#1}}c<{\DC@end}}
\makeatother

\let\orgautoref\autoref

\renewcommand{\autoref}
        {\def\equationautorefname{equation}%
         \def\figureautorefname{Fig.}%
         \def\subfigureautorefname{Fig.}%
         \def\partautorefname{part}%
         \def\chapterautorefname{chapter}%
         \def\sectionautorefname{section}%
         \def\subsectionautorefname{section}%
         \def\subsubsectionautorefname{section}%
         \def\appendixautorefname{appendix}%
         \def\Itemautorefname{item}%
         \def\tableautorefname{Table}%
         \def\lstlistingautorefname{Listing}%
         \orgautoref}
\providecommand{\autorefs}
        {\def\equationautorefname{equations}%
         \def\figureautorefname{Figs.}%
         \def\subfigureautorefname{Figs.}%
         \def\partautorefname{parts}%
         \def\chapterautorefname{chapters}%
         \def\sectionautorefname{sections}%
         \def\subsectionautorefname{sections}%
         \def\subsubsectionautorefname{sections}%
         \def\Itemautorefname{items}%
         \def\tableautorefname{Tables}%
         \def\lstlistingautorefname{Listings}%
         \orgautoref}    
\makeatletter

\newcommand{\psrjna}{PSR\,J0826$+$2637}
\newcommand{\dsunheute}{d_{\odot,today}}
\newcommand{\dsunSN}{d_{\odot,SN}}

\title[The Origin of the Young Pulsar \psrjna{}.]{The Origin of the Young Pulsar \psrjna{} and Its Possible Former Companion HIP~13962.}
\author[Tetzlaff et al.]{N. Tetzlaff$^{1}$\thanks{E-mail:
nina@astro.uni-jena.de}, B. Din\c{c}el$^{1}$, R. Neuh\"auser$^{1}$ and V. V. Kovtyukh$^{2,3}$ \\% M. M. Hohle$^{1}$\\
$^{1}$Astrophysikalisches Institut und Universit\"ats-Sternwarte Jena, Schillerg\"asschen 2-3, 07745 Jena, Germany\\
$^{2}$Astronomical Observatory, Odessa National University, T. G. Shevchenko Park, 65014, Odessa, Ukraine\\
$^{3}$Isaac Newton Institute of Chile, Odessa Branch, Ukraine}

\begin{document}

\date{Accepted. Received; in original form}

\pagerange{\pageref{firstpage}--\pageref{lastpage}} \pubyear{2013}

\maketitle

\label{firstpage}

\begin{abstract} 
We aim to identify the birth place of the young pulsar \psrjna{} in order to determine its kinematic age and give constraints on its radial and spatial (kick) velocity.

Since the majority of neutron star (NS) progenitors are in associations or clusters, we search for a possible origin of the NS inside such stellar groups. We trace back the NS and the centres of possible birth associations and clusters to find close encounters. The kinematic age is then given by the time since the encounter. We use Monte Carlo simulations to account for observational uncertainties and the unknown radial velocity of the NS. We evaluate the outcome statistically. In order to find further indication for our findings, we also search for a runaway star that could be the former companion if it exists.

We find that \psrjna{} was probably born in the small young cluster Stock 7 $\sim\unit[3]{Myr}$ ago. This result is supported by the identification of the former companion candidate HIP~13962 (runaway star with spectral type G0Ia). The scenario predicts a near-zero radial velocity of the pulsar implying an inclination angle of its motion to the line-of-sight of $\unit[87\pm11]{deg}$.

We also present the chemical abundances of HIP 13962. We do not find enhanced $\alpha$ element abundances in the highly evolved star. However, the binary supernova scenario may be supported by the overabundance of r-process elements that could have been ejected during the supernova and accreted by the runaway star. Also, a high rotational velocity of $v\sin i\sim\unit[29]{km/s}$ of HIP 13962 is consistent with evolution in a pre-SN binary system.
\end{abstract}


%%%%%%%%%%%%%%%%%%%%%%%%%%%%%%%%%%%%%%%%%%%%%%%%%%%%%%%%%%%%%%%%%%%%%%%%%%%%%%%%%%%%%%%%%%%%%%%%%%%%%%%%%%%%%%%%%%%%%%%%%%%%%%%%%%%%%%%%%%%%%%%%%

\section{Introduction}

\psrjna{} (PSR\,B0823$+$26) was discovered 45 years ago by \citet{1968IAUC.2100....1C}. Its spin period $P=\unit[0{.}53]{s}$ and period derivative $\dot{P}=\unit[1{.}71\cdot10^{-15}]{s/s}$ \citep{2004MNRAS.353.1311H} yield a characteristic age of $\unit[4{.}92]{Myr}$. It is a middle-age field pulsar that has been detected in X-rays \citep{1993IAUC.5895....2S}. \citet{2004ApJ...615..908B} found that a power law with photon index $\alpha=2{.}5^{+0{.}9}_{-0{.}45}$ already fits well the energy spectrum of \psrjna{}. Adding a blackbody model does not change the fit significantly. So, there is only little thermal contribution from the neutron star (NS). However, \citet{2004ApJ...615..908B} obtain $3\sigma$ upper limits for the temperature of the polar cap ($<\unit[1{.}17\cdot10^6]{K}$) and for the temperature if they assume that the emission comes from the whole surface ($<\unit[0{.}5\cdot10^6]{K}$). Unfortunately, these temperatures do not allow to derive a cooling age of the NS. Considering different cooling models \citep[e.\,g. from][]{2005MNRAS.363..555G,2009AA...496..207P,2010MNRAS.401.2675P} a lower age limit of a few kyr is estimated (assuming a $\unit[1{.}4]{M_\odot}$ NS). \\
Beside the spin-down age the only way to estimate the pulsar's age is to do it kinematically. The knowledge of the birth place then also yields the lifetime of the NS progenitor (hence, its mass) given by the difference between the age of the parent association or cluster and the kinematic age of the NS. Furthermore, if place and time of the SN were known, the radial and spatial (kick) velocity could be constraint. These quantities are not directly measurable. So far, the only way to estimate the radial velocity is to model a bow shock that is created while the NS moves supersonically through the interstellar medium \citep[ISM,][]{2001A&A...380..221V}. Unfortunately, for \psrjna{}, no such bow shock (nor pulsar wind nebula) has been reported yet. If the radial velocity was known, the three-dimensional velocity could be constructed. This is also important to investigate the spin-velocity alignment of pulsars and provides further input to study the kick mechanism in the supernova (SN).

Accepting the spin-down age as a rough estimate (or often an upper limit) of the true age of the NS, \psrjna{} is sufficiently young to trace back its trajectory and identify its birth place. Kinematic ages have been determined for a number of NSs \citep[e.\,g.][]{2001A&A...365...49H,2005A&A...435..625P,2007ApJ...660.1428K,2008AstL...34..686B,2011MNRAS.417..617T,2012PASA...29...98T,2013MNRAS.435..879T,2013MNRAS.429.3517M}. 

Here, we aim to derive the kinematic age of the young pulsar \psrjna{} by identifying its birth place. It is reasonable to assume that NSs are born in young stellar groups such as associations and clusters since the majority of massive stars are observed in associations and clusters ($\gtrsim70$ per cent, e.\,g. \citealt{1998AJ....115..821M,2004ApJS..151..103M}).

\section{Method}\label{sec:method}

We apply the same method as we did in preceding papers \citep{2009MNRAS.400L..99T,2010MNRAS.402.2369T,2011MNRAS.417..617T,2012PASA...29...98T}, so we refer to these publications for details. We construct a few million past trajectories of \psrjna{} and young associations/clusters \citep{2010MNRAS.402.2369T,2012PASA...29...98T,2013PhD...Nina} as well as runaway stars \citep{2011MNRAS.410..190T,2013PhD...Nina} throughout Monte Carlo simulations by varying the observables (parallax, proper motion, radial velocity) within their error intervals. For the radial velocity of \psrjna{}, we assume a reasonable probability distribution derived from the pulsar space velocity distribution \citep{2005MNRAS.360..974H}. From all pairs of trajectories (NS and association/cluster or NS and runaway star), we evaluate the smallest separation $d_{min}$ and the past time $\tau$ at which it occurred. The distribution of separations $d_{min}$ is supposed to obey the distribution of absolute differences of two 3D Gaussians \citep[see e.\,g.][]{2001A&A...365...49H,2012PASA...29...98T}. Since the actual (observed) case is different from the simple model (no 3D Gaussian distributed positions, due to e.\,g. the Gaussian distributed parallax that goes into the position reciprocally, complicated radial velocity distribution, etc.), we adapt the theoretical formulae (equations 1 and 2 in \citealt{2012PASA...29...98T}; here we use the symbols $\mu$ and $\sigma$ for the expectation value and standard deviation, respectively) only to the first part of the $d_{min}$ distribution (up to the peak plus a few more bins, see \citealt{2012PASA...29...98T}). The derived parameter $\mu$ then gives the positional difference between the two objects. 

This procedure was already successfully applied by \citet{2001A&A...365...49H}, \citet{2008AstL...34..686B,2009AstL...35..396B} and us \citep{2009MNRAS.400L..99T,2010MNRAS.402.2369T,2011MNRAS.417..617T,2012PASA...29...98T,2013MNRAS.435..879T}. We performed an investigation of (artificial) test cases that showed that it is well possible to recover place and time of the formation of a NS. If the former companion to the NS progenitor (if it exists) could be identified, the kinematic age is well consistent with the true NS age for 90 per cent of the test cases \citep{2013PhD...Nina}. While the rate of identifying the birth association or cluster of a nearby NS is 70 per cent, the former companion could be identified for 35 per cent of the test cases.\\

For \psrjna{}, we adopt the following parameters for the right ascension $\alpha$, declination $\delta$ (J2000), parallax $\pi$ and proper motion $\left(\mu_{\alpha}^*,\mu_\delta\right)$:
\begin{equation}
	\begin{array}{l c l}
	\alpha &=& 08^\mathrm{h}26^\mathrm{m}51^\mathrm{s}\hspace{-0.8ex}.3833,\ \delta\ =\ +26^\circ37\mathrm{'}23\mathrm{''}\hspace{-0.8ex}.79 \\
	       && \mbox{\citep{2004MNRAS.353.1311H}},\\
	\pi &=& \unit[8{.}16\pm0{.}80]{mas}\ \mbox{\citep{1986AJ.....91..338G}},\\
	\mu_{\alpha}^* &=& \unit[325{.}9\pm2{.}3]{mas/yr},\\
	\mu_{\delta} &=& \unit[-59{.}2\pm2{.}1]{mas/yr} \\
	              &&\mbox{\citep{1982MNRAS.201..503L}},
	\end{array}\label{eq:0826input}
\end{equation}
\noindent where $\mu_{\alpha}^*$ is the proper motion in right ascension corrected for declination.\\

Positional and kinematic data of young Hipparcos runaway stars are taken from \citet{2011MNRAS.410..190T} (see also \citealt{2013PhD...Nina}).

%______________________________________________________________

\section{Results}\label{sec:results}

By tracing back the past trajectories of \psrjna{} and a large sample of young associations and clusters (see \citealt{2010MNRAS.402.2369T,2011MNRAS.417..617T,2012PASA...29...98T}), seven associations and clusters were found for which the NS could have been inside their boundaries during the past $\unit[5]{Myr}$. These seven associations/clusters are therefore candidates to have hosted the birth place of \psrjna{}. The NS either originated in a nearby ($<\unit[200]{pc}$) association $\lesssim\unit[1]{Myr}$ ago or in a more distant one ($\sim\unit[600-900]{pc}$) up to $\sim\unit[5]{Myr}$ ago (\autoref{tab:0826_ass}). In the latter cases a near-zero radial velocity of the NS is necessary. Two results are given in \autoref{tab:0826_ass} for these cases: ($^*$) using a probability distribution for the NS radial velocity according to \citet{2005MNRAS.360..974H} and ($^\#$) using a uniform distribution in the range of $-1500$ to $\unit[+1500]{km/s}$.
\begin{table*}
\centering
\caption{Present-day parameters and encounter position and time for possible parent associations/clusters of \psrjna{} (Column 1). \newline
Column 2: predicted separation between the encounter and the centre of the association (expectation value $\mu$ and standard deviation $\sigma$ inferred from equations 1 or 2 in \citealt{2012PASA...29...98T}, but peak value and 68 per cent error interval for Cam OB1 because the equation was not adaptable to the $d_{min}$ histogram). Column 3: encounter time $\tau$. Columns 4-8: Predicted present NS parameters (heliocentric radial velocity $v_r$, proper motion $\mu_\alpha^*$ and $\mu_\delta$, peculiar space velocity $v_{sp}$, parallax $\pi$). %Note that it is possible that the derived value for $v_r$ is larger than that of $v_{sp}$ because $v_r$ is heliocentric whereas $v_{sp}$ is the peculiar velocity of the NS that reflects its kick velocity. 
Columns 9-12: Predicted SN position (SN distance at the time of the SN ($\dsunSN$) and as seen from Earth today ($\dsunheute$) and Galactic coordinates, $l$ and $b$, J2000.0, as seen from Earth today). Error bars denote $\unit[68]{\%}$ confidence (for the derivation of the parameters we refer to \citealt{2010MNRAS.402.2369T}.). Column 13: estimated progenitor mass $M_{prog}$ derived from the progenitor lifetime (age of the parent association/cluster, minus the time $\tau$ since the potential SN) using evolutionary models from \citet{1980FCPh....5..287T,1989A&A...210..155M,1997PhDT........31K}. Ages of associations/clusters: $\unit[7-14]{Myr}$ (Cam OB1, \citealt{1985A&A...143L...7S}), $\unit[50-71]{Myr}$ ($\alpha$ Per, e.\,g. \citealt{1999AJ....117..354D,2001AstL...27..386L}), $\unit[50-71]{Myr}$ (Cas-Tau, \citealt{1999AJ....117..354D}), $\unit[12-45]{Myr}$ (Stock 7, \citealt{2001A&AT...20..607L,2005A&A...438.1163K,2010MNRAS.407.2109V}; \L. Bukowiecki, priv. comm.), $\unit[56]{Myr}$ (NGC 433, \citealt{2012AcA....62..281B}), $\unit[71]{Myr}$ (NGC 1027, \citealt{2012AcA....62..281B}), $\unit[79]{Myr}$ (NGC 1444, \citealt{2012AcA....62..281B}).}\label{tab:0826_ass}
\setlength\extrarowheight{4pt}
\fontsize{7}{8.4} \selectfont %2. Wert = 1.2 x 1. Wert
\begin{tabular}{@{}c >{$}c<{$}@{} >{$\hspace*{1em}}r<{$}@{} >{$\hspace*{1em}}r<{$}@{} o{4.2} o{4.2} >{$\hspace*{1em}}r<{$}@{} >{$\hspace*{1em}}r<{$}@{} >{$\hspace*{1em}}c<{$}@{} >{$\hspace*{1em}}c<{$}@{} >{$\hspace*{1em}}r<{$}@{} >{$\hspace*{1em}}r<{$}@{} >{$\hspace*{1em}}c<{$}@{}}
\toprule
Assoc.	&	\multicolumn{1}{c}{$\mu,\sigma$} & \multicolumn{1}{c}{$\tau$} &	\multicolumn{5}{c}{Predicted present-day NS parameters}	&	\multicolumn{4}{c}{Predicted SN/SNR position} & M_{prog}\\ 
	&\multicolumn{1}{c}{}	&  &	\multicolumn{1}{c}{$v_r$}			& \multicolumn{1}{c}{$\mu_{\alpha}^*$} & \multicolumn{1}{c}{$\mu_{\delta}$} & \multicolumn{1}{c}{$v_{sp}$} &	\multicolumn{1}{c}{$\pi$}			&	\dsunSN & \dsunheute 	& \multicolumn{1}{c}{$l$}	& \multicolumn{1}{c}{$b$} & \\ 
		& \multicolumn{1}{c}{[pc]} & \multicolumn{1}{c}{[Myr]}	& \multicolumn{1}{c}{[km/s]} & \multicolumn{1}{c}{[mas/yr]} & \multicolumn{1}{c}{[mas/yr]} & \multicolumn{1}{c}{[km/s]} & \multicolumn{1}{c}{[mas]} & \multicolumn{1}{c}{[pc]} & \multicolumn{1}{c}{[pc]} & \multicolumn{1}{c}{[$^\circ$]} & \multicolumn{1}{c}{[$^\circ$]} & \mathrm{[M_\odot]}\\\midrule
Cam OB1	& \multicolumn{1}{c}{$62^{+92}_{-35}$} &3{.}56^{0{.}80}_{0{.}81}	& -13^{+29}_{-37}	&	61{.}1+3{.}0	&	-90{.}0+2{.}0	&	173^{+56}_{-28}	&	2{.}6^{+0{.}7}_{-0{.}6}	&	827^{+65}_{-39}	& 776^{+57}_{-49} &	144{.}0^{+5{.}8}_{-2{.}9}	&	1{.}1^{+2{.}8}_{-3{.}4} & \gtrsim15 \\
$\alpha$ Per (Per OB3)	& 17{.}8,7{.}2 & 0{.}75^{+0{.}18}_{-0{.}15}	& 315^{+112}_{-43}	&	61{.}3+2{.}9	&	-90{.}0+2{.}0	&	363^{+111}_{-45}	&	2{.}7^{+0{.}3}_{-0{.}3}	&	159^{+23}_{-8} & 157^{+21}_{-9}	&	141{.}1^{+1{.}0}_{-0{.}8}	&	-2{.}0^{+1{.}1}_{-1{.}1} & 5-7\\
Cas-Tau	& 68{.}2,6{.}7 & 0{.}69^{+0{.}20}_{-0{.}12}	& 367^{+102}_{-65}	&	61{.}4+2{.}9	&	-89{.}9+2{.}0	&	422^{+92}_{-77}	&	2{.}7^{+0{.}4}_{-0{.}4}	&	141^{+28}_{-17} & 142^{+28}_{-17}	&	141{.}1^{+1{.}5}_{-1{.}3}	&	-2{.}3^{+1{.}7}_{-1{.}3} & 6-7 \\
Stock 7$^\mathsf{a,*}$	& 0,31{.}9 & 2{.}1^{+0{.}9}_{-0{.}6}	& -10^{+157}_{-102}	&	59{.}2+2{.}1	&	-90{.}5+2{.}0	&	243^{+114}_{-99}	&	1{.}9^{+0{.}4}_{-0{.}6}	&	704^{+15}_{-16}	& 680^{+12}_{-20} &	139{.}3^{+1{.}3}_{-2{.}0}	&	-0{.}1^{+0{.}9}_{-0{.}2} & 6-31\\
Stock 7$^\mathsf{\#}$	& 0,34{.}5 & 3{.}1^{+0{.}7}_{-0{.}7}	& -12^{+52}_{-9}	&	59{.}6+2{.}1	&	-90{.}4+2{.}0	&	187^{+55}_{-35}	&	2{.}7^{+0{.}5}_{-0{.}7}	&	695^{+28}_{-16}	& 661^{+33}_{-7} &	141{.}7^{+1{.}8}_{-2{.}5}	&	0{.}4^{+0{.}2}_{-1{.}1} & 6-30\\
NGC 433$^\mathsf{a,*}$ &  0,227{.}7 & 3{.}2^{+1{.}0}_{-1{.}0}	& 38^{+77}_{-71}	&	57{.}5+1{.}9	&	-91{.}1+2{.}0	&	518^{+102}_{-106}	&	1{.}0^{+0{.}1}_{-0{.}2}	&	1826^{+78}_{-5}	& 1747^{+57}_{-10} &	128{.}4^{+1{.}9}_{-1{.}4}	&	-3{.}9^{+0{.}8}_{-0{.}4} & \gtrsim6^\mathrm{b}\\
NGC 433$^\mathsf{\#}$ &  0,196{.}6 & \sim4{.}7	& -7^{+34}_{-26}	&	58{.}8+2{.}7	&	-90{.}7+1{.}9	&	394^{+29}_{-34}	&	1{.}3^{+0{.}2}_{-0{.}1}	&	1934^{+42}_{-41}	& 1840^{+43}_{-50} &	128{.}8^{+2{.}4}_{-0{.}8}	&	-2{.}3^{+1{.}2}_{-1{.}5} & \gtrsim6^\mathrm{b}\\
NGC 1027$^\mathsf{a,*}$ & 0,26{.}5	& 2{.}1^{+0{.}9}_{-0{.}6}	& 17^{+97}_{-43}	&	59{.}1+1{.}7	&	-90{.}9+1{.}8	&	213^{+104}_{-59}	&	2{.}0^{+0{.}5}_{-0{.}5}	&	687^{+8}_{-12}	& 653^{+22}_{-9} &	139{.}4^{+0{.}5}_{-1{.}3}	&	-0{.}2^{+1{.}1}_{-0{.}5} & \gtrsim5^\mathrm{b}\\
NGC 1027$^\mathsf{\#}$ & 0,36{.}2	& \gtrsim1{.}5	& -36^{+17}_{-5}	&	61{.}0+2{.}0	&	-90{.}0+2{.}0	&	234^{+52}_{-12}	&	2{.}2^{+0{.}2}_{-0{.}4}	&	1335^{+16}_{-6}	& 1251^{+29}_{-19} &	138{.}9^{+0{.}9}_{-0{.}6}	&	-0{.}3^{+0{.}4}_{-0{.}8} & \gtrsim5^\mathrm{b}\\
NGC 1444$^\mathsf{a,*}$ & 0,252{.}4 & \gtrsim3{.}8	& -59^{+5}_{-5}	&	63{.}8+2{.}0	&	-89{.}1+1{.}8	&	173^{+26}_{-9}	&	2{.}7^{+0{.}4}_{-0{.}1}	&	1069^{+21}_{-32}	& 973^{+32}_{-20} &	148{.}7^{+0{.}7}_{-0{.}8}	&	3{.}3^{+0{.}4}_{-0{.}9} & \gtrsim5^\mathrm{b}\\
NGC 1444$^\mathsf{\#}$ & 0,178{.}9 & \sim4{.}1	& -82^{+37}_{-8}	&	61{.}1+2{.}7	&	-90{.}0+2{.}0	&	187^{+26}_{-8}	&	2{.}5^{+0{.}4}_{-0{.}2}	&	1055^{+46}_{-28}	& 997^{+52}_{-41} &	144{.}5^{+6{.}1}_{-3{.}6}	&	4{.}8^{+4{.}1}_{-1{.}9} & \gtrsim5^\mathrm{b}\\
\bottomrule
\multicolumn{13}{p{0.97\textwidth}}{\scriptsize $^\mathrm{a}$Since small $v_r$ of the NS were predicted using a $v_r$ distribution consistent with the spatial velocity distribution of pulsars by \citet{2005MNRAS.360..974H} (cases $^*$), the calculations were repeated using a uniform $v_r$ distribution in the range $\unit[1500-+1500]{km/s}$ (cases $^\#$). The results did not change significantly (except the SN distance for NGC 1027).}\\
\multicolumn{13}{p{0.97\textwidth}}{\scriptsize $^\mathrm{b}$These $M_{prog}$ estimates are smaller than the minimum mass of a star that can experience a core-collapse SN ($\sim\unit[8-9]{M_\odot}$, \citealt[e.\,g.][]{2003ApJ...591..288H}). However, isochrone ages of NGC 433, NGC 1027 and NGC 1444 were taken from \citet{2012AcA....62..281B} and are upper limits (\L. Bukowiecki, priv. comm.). Then, higher progenitor masses ($>\unit[8]{M_\odot}$) can be derived.}
\end{tabular}
\end{table*}
Considering the estimated masses of the SN progenitor star given in that table, it is less likely that one of the nearby associations Per OB3 ($\alpha$ Per) and Cas-Tau is the parent association of \psrjna{} since the predicted masses are smaller than the minimum mass of a star that can experience a core-collapse SN ($\sim\unit[8-9]{M_\odot}$, \citealt[e.\,g.][]{2003ApJ...591..288H}). We note that \citet{2001A&A...365...49H} also suggested Per OB3 being the birth association of \psrjna{} with an age of $\sim\unit[1]{Myr}$, consistent with our result. Their estimated pulsar radial velocity is somewhat smaller ($\sim\unit[100]{km/s}$), but for a slightly larger kinematic age ($\sim\unit[1]{Myr}$), we achieve the same estimate. However, we consider it more probable that \psrjna{} was born $\sim\unit[2-4]{Myr}$ ago in an association or cluster with a distance of $\gtrsim\unit[700]{pc}$ in the Camelopardalis region.\\

\begin{figure}
\centering
\includegraphics[width=0.45\textwidth, viewport= 20 210 580 625]{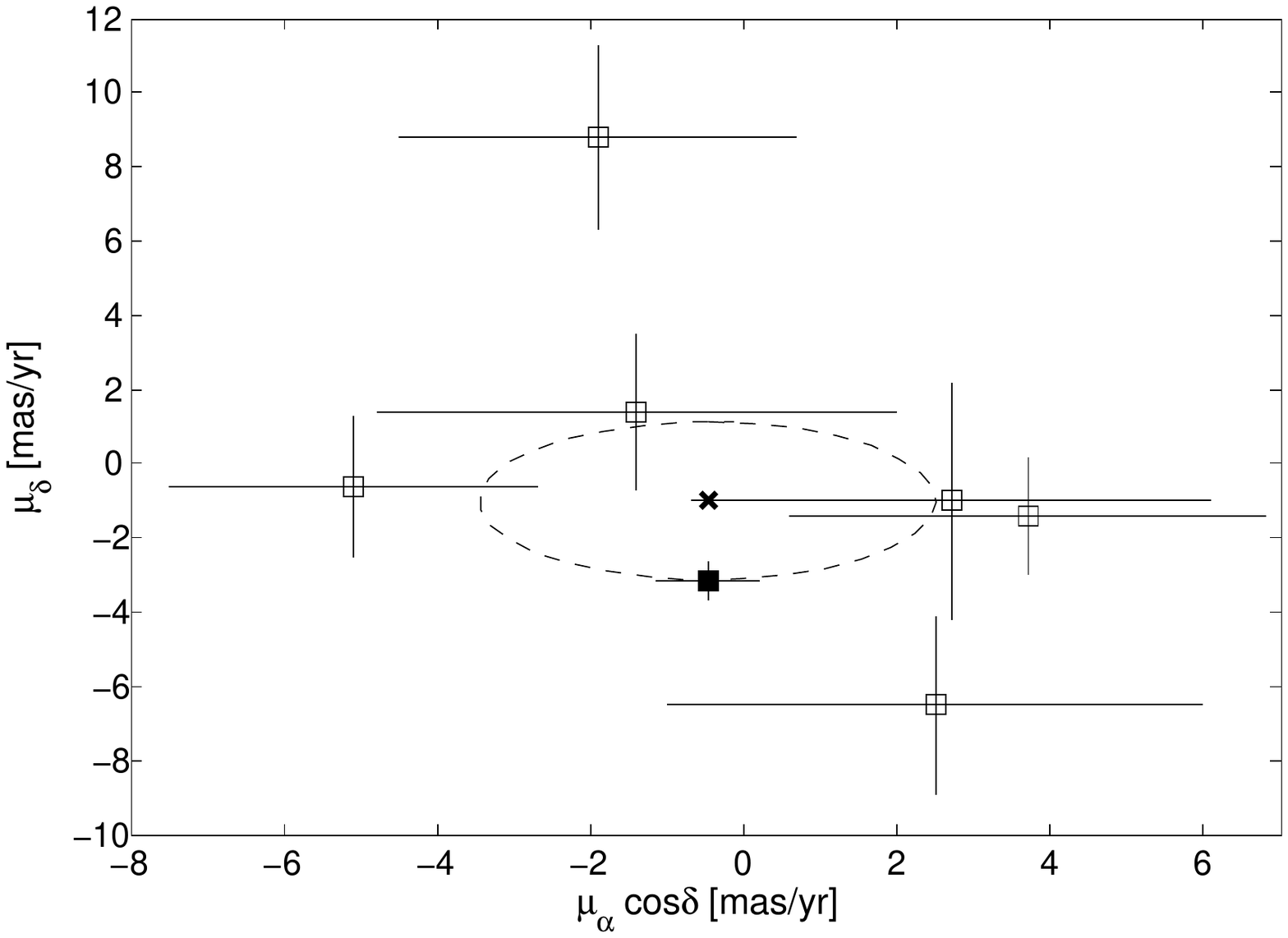}
\caption{Proper motions of stars that are proposed member stars (rectangles, \citealt{1998A&A...335L..65H,2007AA...474..653V}) of a cluster around HIP 13962 (filled rectangle, \citealt{2007AA...474..653V}), \citet{2009AN....330..807T}. The median value is marked as a cross. The dashed ellipse represents the median deviation from this value. The dispersion is more than $4-5$ times larger than the typical value for clusters ($\sim\unit[5]{km/s}$ in each direction, \citealt{2010MNRAS.402.2369T}, corresponding to $\sim\unit[0{.}6]{mas/yr}$ at a cluster distance of $\unit[1{.}7]{kpc}$, \citealt{2009AN....330..807T}).}\label{fig:Turnerpm}
\end{figure}

\begin{figure}
\centering
\includegraphics[width=0.45\textwidth, viewport= 20 210 580 625]{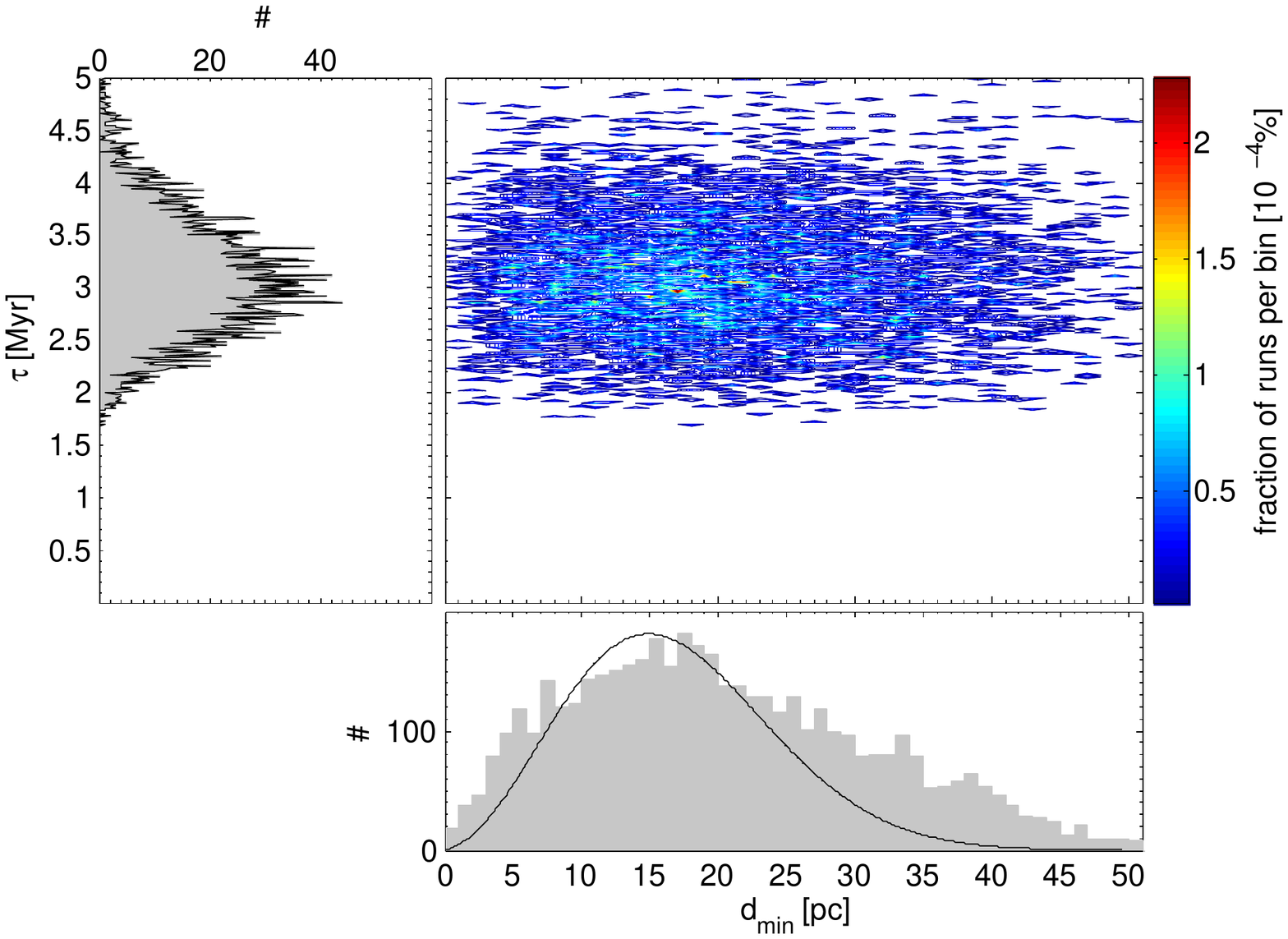}
\caption[$d_{min}$ and $\tau_{kin}$ distributions for \psrjna{} and HIP 13962.]{Distributions of minimum separations $d_{min}$ and corresponding flight times $\tau$ for encounters between \psrjna{} and HIP 13962, only those Monte Carlo runs are shown for which both stars were within $\unit[30]{pc}$ from the centre of Stock 7. The solid curves drawn in the $d_{min}$ histograms (bottom panel) represent the theoretically expected distribution (see \autoref{sec:method}) with $\mu=0$ and $\sigma=\unit[13{.}0]{pc}$, adapted to the first part of the histogram.}\label{fig:0826_13962}
\end{figure}

\begin{table}
\centering
\caption[Predicted current parameters of \psrjna{} and SN position and time.]{Predicted current parameters of \psrjna{} and SN position and time if HIP 13962 was the former companion.}\label{tab:0826predparmain}
{\footnotesize
\begin{tabular}{l  c}
\toprule
\multicolumn{2}{p{0.35\textwidth}<{\centering}}{Predicted present-day parameters of \psrjna{}}\\\midrule
$v_r$ [km/s]             	& $1^{+36}_{-15}$	\\
$\pi$ [mas]       	& $2{.}6^{+0{.}5}_{-0{.}5}$\\
$\mu_{\alpha}^*$ [mas/yr]  & $60{.}0\pm2{.}4$	\\
$\mu_\delta$ [mas/yr]      & $-90{.}4\pm1{.}9$	\\
$v_{sp}$ [km/s]					 		& $183^{+39}_{-32}$			\\\midrule
\multicolumn{2}{p{0.35\textwidth}<{\centering}}{Predicted SN/SNR pos. and time}\\\midrule
$\dsunSN$ [pc] & $707^{+19}_{-32}$\\
$\dsunheute$ [pc] & $671^{+23}_{-25}$ \\
$l$ [$^\circ$]     & $140{.}5^{+1{.}6}_{-1{.}2}$	\\
$b$ [$^\circ$]     & $-0{.}2^{+0{.}7}_{-1{.}4}$	\\
$\tau$ [Myr]   	& $3{.}0\pm0{.}6$	\\\bottomrule
\multicolumn{2}{p{0.35\textwidth}}{\scriptsize The parameter designations are as in \autoref{tab:0826_ass}. Since for an origin in Stock 7, the NS $v_r$ was found to be very small, the results given here were obtained using a NS $v_r=\unit[0\pm100]{km/s}$ in the Monte Carlo simulations.}
\end{tabular}
}
\end{table}
To find further evidence for a particular birth place of \psrjna{}, we check whether any runaway star (from \citealt{2011MNRAS.410..190T,2013PhD...Nina}) could have been at the same place at the same time as the NS. \\
Among 137 runaway stars that could have come close to the NS sometime in the past $\unit[5]{Myr}$, only the encounter between \psrjna{} and the runaway star HIP 13962 could have happened inside a possible birth association/cluster of \psrjna{}. We also considered whether the SN that ejected both stars occurred outside any association or cluster by evaluating the significance of the encounters. We compare the probability of each possible encounter with a reference probability of an encounter between a randomly chosen NS and a randomly chosen runaway star (the idea was originally developed by \citealt{2010AstL...36..116C} and adapted to our work, for details we refer to \citealt{2012PASA...29...98T} and \citealt{2013PhD...Nina}). \\
With this method, we did not find evidence for an isolated SN that ejected \psrjna{} and a runaway star, whereas the identification of the G0Ia type \citep{2009AN....330..807T} runaway star HIP 13962 as a former companion candidate to \psrjna{} supports an origin of both stars in the small cluster Stock 7. The potential SN by which both the runaway star and the NS were ejected, was located well inside the cluster boundaries. Considering Monte Carlo runs that yield separations between HIP 13962 and \psrjna{} that are smaller than a few parsecs (the smallest separation found is $\unit[0{.}2]{pc}$) results in distributions of separations between the NS and the cluster centre and the runaway star and the cluster centre, respectively, that are consistent with that the SN event took place at the centre of Stock 7, i.\,e. $\mu=0$ in both cases. The standard deviation is $\sigma=\unit[31]{pc}$ (in both cases) and consistent with the error corrected radius of the cluster\footnote{Due to the uncertainties of the kinematic properties of the cluster, the apparent radius increases as the (calculation) time increases to the past.\label{footn:Rcorr}. The nominal radius of the cluster is $\sim\unit[2]{pc}$, \citealt{2005A&A...440..403K} whereas the error corrected radius $R_{corr}=\unit[30]{pc}$.}. The distribution of separations $d_{min}$ between the NS and the runaway star is also consistent with $\mu=0$, i.\,e. they were ejected the same SN event $\unit[3{.}0\pm0{.}6]{Myr}$ ago (\autoref{fig:0826_13962}). This kinematic age is comparable to and not larger than the characteristic age of \psrjna{}, $\tau_{char}=\unit[4{.}92]{Myr}$ \cite{2004MNRAS.353.1311H}. The present NS parameters and position of the predicted SN are given in \autoref{tab:0826predparmain}. \\
\cite{2009AN....330..807T} propose that HIP 13962 is a member of a previously unknown sparsely populated young cluster (with an age of $\unit[9\pm1]{Myr}$) that is presently dissolving into the field. The existence of that cluster is questionable, however. If existing, the cluster is only marginally detectable as cluster (see Figure 9 in \citealt{2009AN....330..807T}). It is not detectable in infrared data (2MASS $\mathrm{JHK_S}$ \cite{2003yCat.2246....0C}, \L. Bukowiecki, priv. comm., see also \cite{2012AcA....62..281B} for the method of detection). Moreover, the nine potential member stars listed by \citet{2009AN....330..807T} do not share a common proper motion (\autoref{fig:Turnerpm}). Hence, the existence of this cluster is arguable.

The evolutionary age of HIP 13962 as post-main sequence star is $\unit[20\pm4]{Myr}$ \citep{2011MNRAS.410..190T}, in agreement with the age of its proposed parent cluster Stock 7 ($\unit[13-16]{Myr}$ from isochrone fitting, \citealt{2001A&AT...20..607L,2005A&A...438.1163K}; $\unit[16\pm4]{Myr}$ from kinematics, \citealt{2010MNRAS.407.2109V}) and significantly larger than the proposed age of Turner's cluster. The age of the cluster Stock 7 could be as high as $\unit[45]{Myr}$ (\L. Bukowiecki, priv. comm., from isochrone fitting, see also \cite{2012AcA....62..281B} for the method). Then, the possible former companion to \psrjna{} would be a blue straggler as expected for runaway stars that were ejected during a SN in a former binary system \citep{2001A&A...365...49H}. \\
Using an age range of Stock 7 of $\unit[12-45]{Myr}$, the estimated mass of the NS progenitor is $m_1\lesssim\unit[30]{M_\odot}$. %Taking into account that the proposed former companion HIP 13962 %has an age of $\unit[20\pm4]{Myr}$ and 
%evolved $\sim\unit[3]{Myr}$ after the SN and that it might be a blue straggler (i.\,e. we assume an age for the runaway star of $\unit[12-45]{Myr}$), $M_{prog}$ can be restricted to $\unit[8-28]{M_\odot}$ from the progenitor lifetime \citep{1980FCPh....5..287T,1989A&A...210..155M,1997PhDT........31K}, corresponding to a spectral type of xxx on the main sequence \citep{Schmidt-Kaler1982,2010AN....331..349H}. 
Further restriction arises from that the SN progenitor must be more massive than its former companion. The mass of HIP 13962 is $\sim\unit[12]{M_\odot}$ \citep{2011MNRAS.410..190T}. On the main sequence, this star should have had a comparable mass, since in this mass regime the mass loss is small ($\sim$\,5 per cent, \citealt[e.\,g.][]{1992A&AS...96..269S}). Hence, the NS progenitor had a mass between 12 and $\sim\unit[30]{M_\odot}$. This is consistent with that its mass must be higher than the mass of the earliest present member of the parent cluster Stock 7 (HD 15239, B2.5:Vn shell, \citealt{1973A&AS...11....3M,1980PASP...92...60A}, hence mass of $\sim\unit[7-8]{M_\odot}$, \citealt{Schmidt-Kaler1982,2010AN....331..349H}). The mass ratio of the former binary system is then $m_2/m_1 > 0{.}4$. This is consistent with observations of massive binaries \citep[e.\,g.][]{2007ApJ...670..747K,2008MNRAS.386..447S}.
%\clearpage
%

The present parameters of \psrjna{} as well as the time and position of the SN if HIP 13962 is the former companion to the NS are given in \autoref{tab:0826predparmain}.\\

If HIP 13962 experienced a nearby SN explosion, the photosphere of the star might be polluted by $\alpha$ process elements which are ejected by SNe into the ISM at high rates. $\alpha$ enhancement was discovered in a few sources like the hypervelocity star HD 271791 \citep{2008ApJ...684L.103P} and the optical companions of two black hole binary systems; Nova Scorpii 1994 \citep{1999Natur.401..142I} and V404 Cyg \citep{2011ApJ...738...95G}. O, Ne, Mg, Si, S and Ca are expected to be overabundant in the photosphere compared to iron (for SN/hypernova yields, see \citealt{2006NuPhA.777..424N}). Hence, we investigated the chemical abundances of HIP 13962.

\begin{figure*}
\centering
\hspace*{-2em}\includegraphics[width=0.4\textwidth, viewport=40 44 550 425]{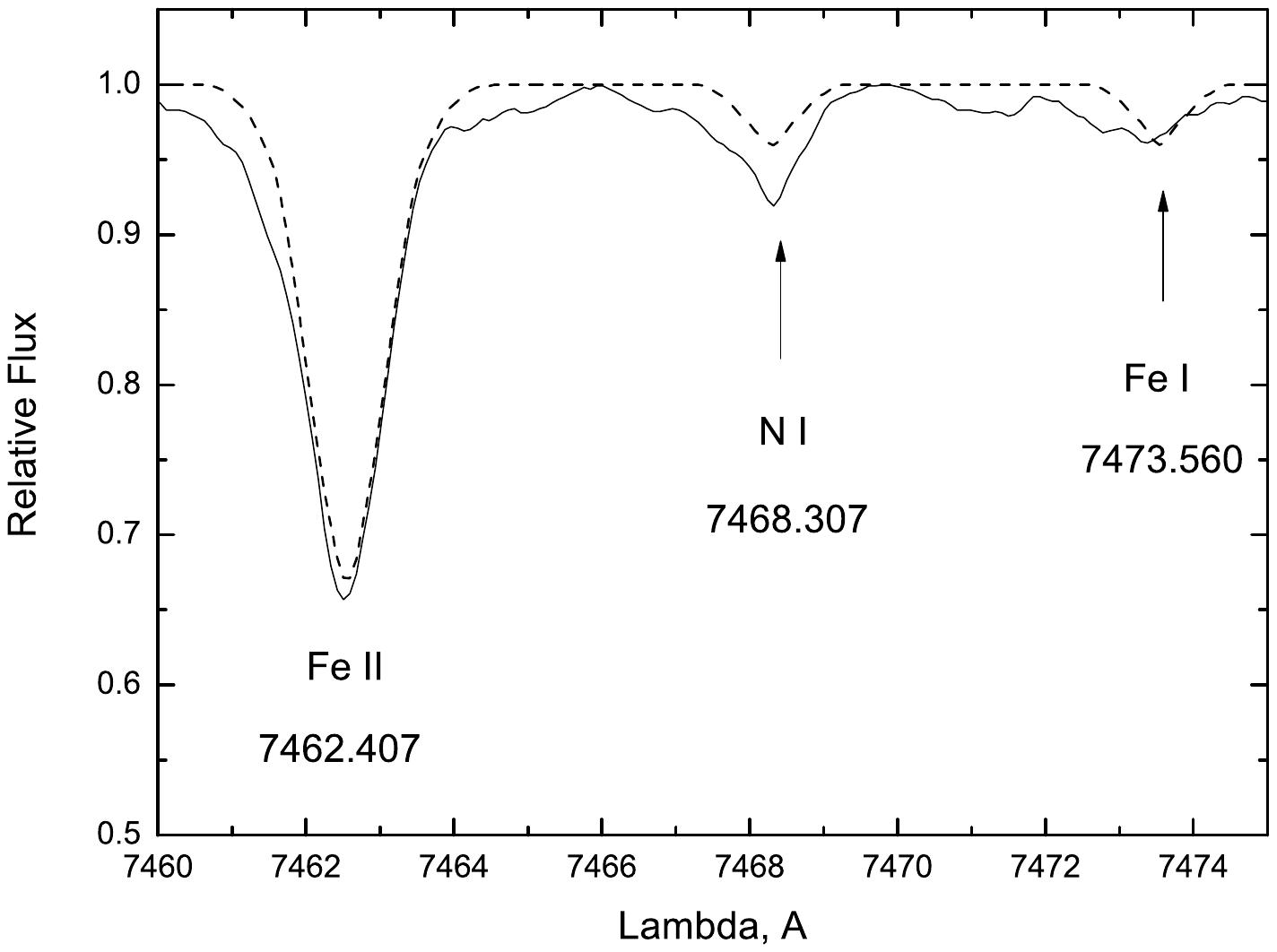}\hspace*{-4em}\nolinebreak
\includegraphics[width=0.4\textwidth, viewport=40 44 550 425]{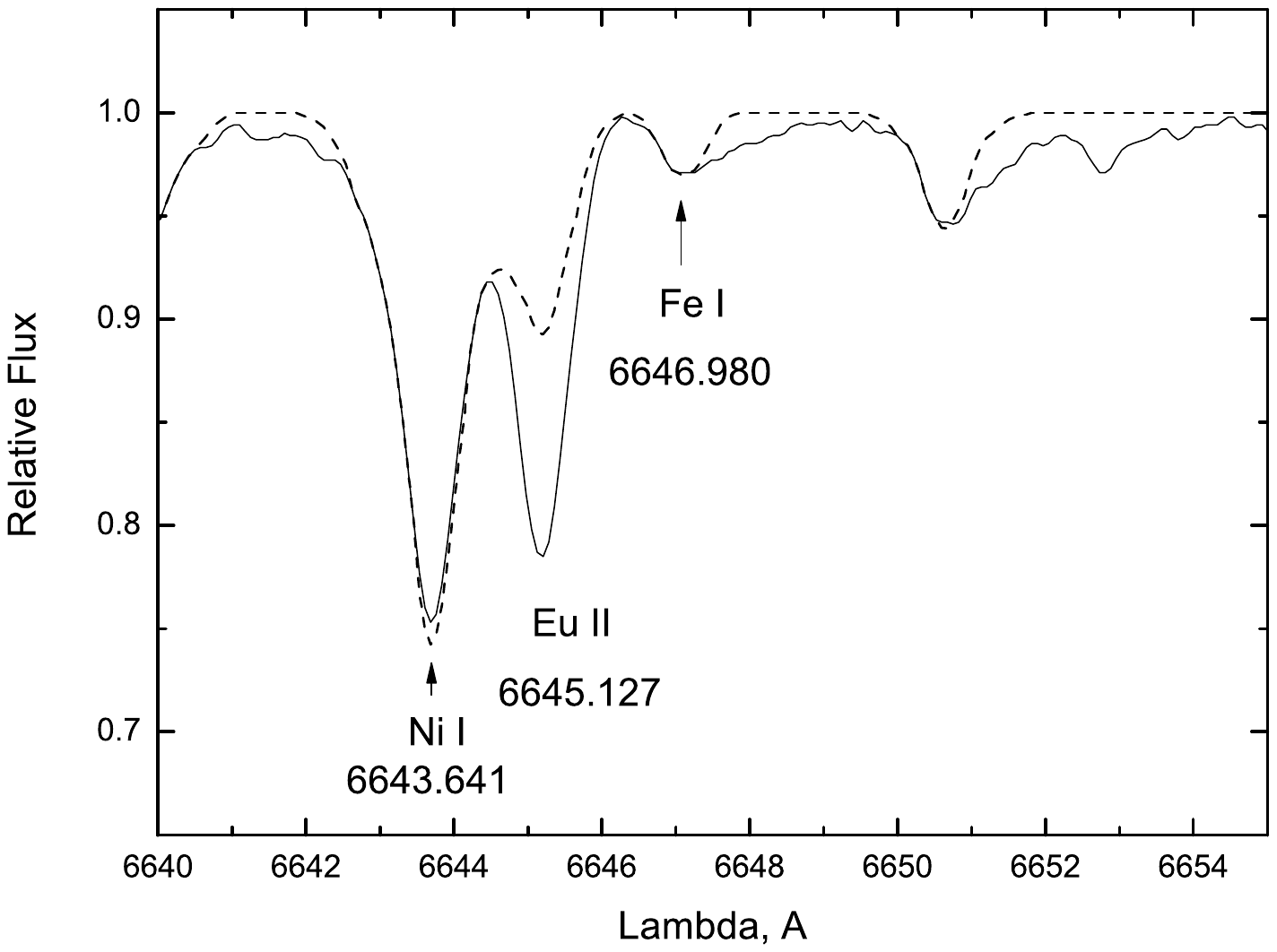}\hspace*{-4em}\nolinebreak
\includegraphics[width=0.4\textwidth, viewport=40 44 550 425]{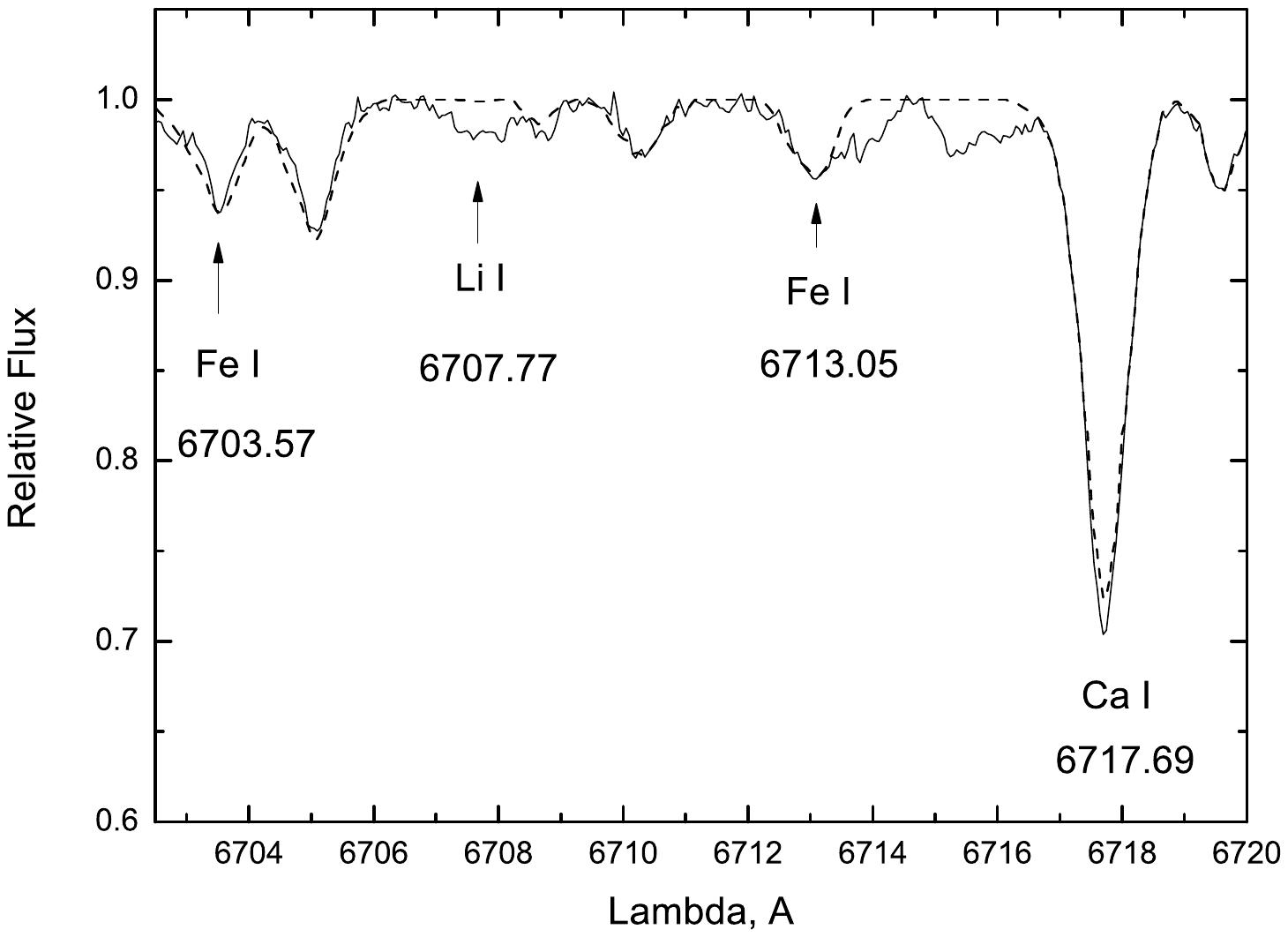}
\caption{Spectrum of HIP 13962 (solid line) showing the regions of the Nitrogen 7468\,\AA{} (left panel), Europium 6645\,\AA{} (middle panel) and Lithium 6708\,\AA{} (right panel) lines. The dashed lines are synthetic spectra, computed with solar abundance.}\label{fig:N_Li}
%\caption{Spectrum of HIP 13962 (solid line) showing the Nitrogen 7468\,\AA{} line region. The dashed line represents a synthetic spectrum, computed with [N/H]\,=\,0.}\label{fig:N_Li}
\end{figure*}

\begin{figure}
\centering
\includegraphics*[width=0.45\textwidth, viewport= 0 270 708 750]{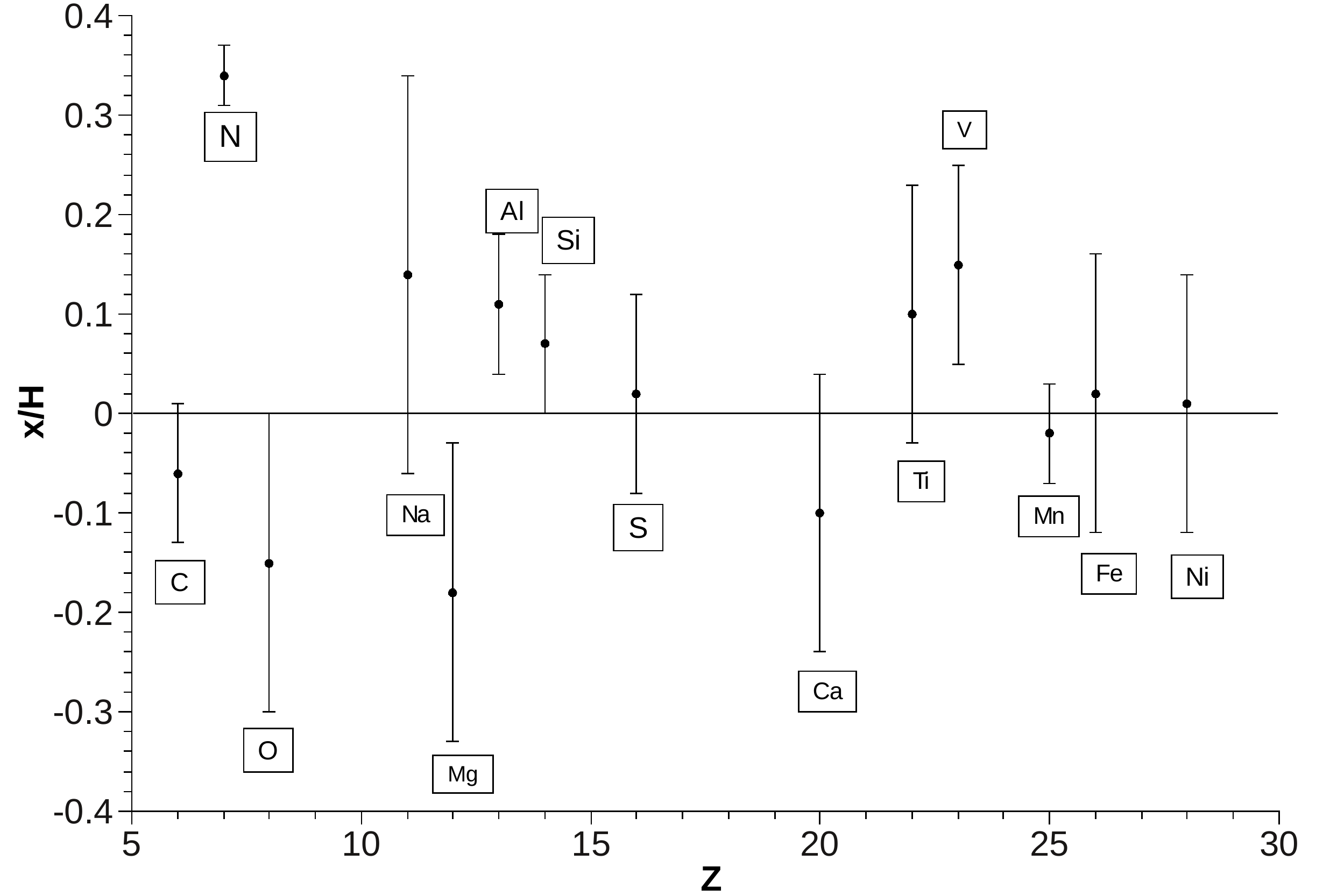}
\caption{Abundances of $\alpha$ process elements for HIP\,13692 relative to solar values. For comparison, abundances of iron peak elements are also shown. Additional significant overabundances: $\mathrm{[Eu/H]}=0{.}39\pm0{.}20$, $\mathrm{[Gd/H]}=0{.}42\pm0{.}20$, $\mathrm{[Li/H]}=0{.}89\pm0{.}20$.}\label{fig:abundances}
\end{figure}

Based on archival ELODIE (Observatoire de Haute-Provence, resolution $R\sim42\,000$) data, the stellar parameters of HIP 13962 were derived \citep{2007MNRAS.378..617K,2012MNRAS.423.3268K}, effective temperature $T_{eff}=\unit[5871\pm130]{K}$ \citep{2007MNRAS.378..617K}, surface gravity $\log g=1{.}2$, micro turbulence $\xi=\unit[11{.}5]{km/s}$ \citep{2012MNRAS.423.3268K} and projected rotational velocity $v\sin i\sim\unit[90]{km/s}$ \citep{2009AN....330..807T}. We determine $v\sin i=\unit[29{.}0\pm2{.}3]{km/s}$. This value is still significantly larger than the typical $v\sin i$ for FG supergiants ($\unit[5-12]{km/s}$, \citealt{2002A&A...395...97D}). HIP 13962 is a yellow supergiant at nearly solar metallicity. The abundances for each element relative to hydrogen are given in \autoref{tab:abundance}.\\
The source has an unexpectedly high Li abundance (\autoref{fig:N_Li}). This can be due to the high rotational velocity \citep{2005KFNT...21..141K} and/or a relatively young age, if the star was a pre-main sequence star (instead of post-main sequence star) -- consistent with a recent core-collapse SN in its system. The fast rotation can be due to the accretion from the primary during pre-SN binary evolution. The Li enrichment is also seen in the case of V404 Cyg \citep{1992Natur.358..129M}. The most plausible mechanism is thought to be the neutron induced spallation of CNO elements on the companion atmosphere \citep{1999ApJ...512..332G}. In our case, we suggest energetic protons and $\alpha$ particles from the SN as the source of the spallation of CNO elements. So, Li enhancement might be explained by the spallation on the atmosphere of HIP 13962 due to the SN event. \\
While it is hard to comment on the whole picture because of the high errors, the Nitrogen enrichment seems clear (\autorefs{fig:N_Li}, \ref{fig:abundances}). However, it is expected in supergiants together with the underabundance of O, C according to convective dredge up. The low abundance of Mg and a higher abundance of Na are also observed in other supergiants \citep{2008AJ....136...98L}. S and Si abundances are consistent with that of iron. So, there is no clear $\alpha$ enhancement in HIP 13962. N enrichment is due to the dredge-up, but it can also arise from mass accretion from the primary in the pre-SN binary system. \\
Furthermore, the accretion from the expanding material of the SN also depends on the binary separation; $M_{SN}=M_{ejct}\cdot R^{2}_{sec}/4a^{2}$, where $M_{SN}$ is the geometric impacting mass, $R_{sec}$ is the radius of the secondary and $a$ is the binary separation \citep{1981ApJ...243..994F}. E.\,g. for a twin binary with masses $\unit[12+12]{M_\odot}$ with binary separation $\gtrsim\unit[800]{R_\odot}$, the accreted mass is $\lesssim\unit[4\cdot10^{-4}]{M_\odot}$ for an ejected mass of $\unit[10]{M_\odot}$ and a secondary radius of $\unit[10]{R_\odot}$. This is probably too low to be detectable. However, the mechanism is not very well understood and searching for $\alpha$ enhancement is still important. \\
Another point is that the photosphere is rich in rare-earth elements (\autoref{tab:abundance}).
Among these features, Europium and Gadolinium are mostly r-process elements in solar composition \citep{2000ApJ...544..302B}.
Compared to the mean chemical abundances of 64 F to M type supergiants given in \citep{1989ApJS...71..559L}, HIP 13692 is slightly enriched in Eu.
The nearby SN might be responsible for the r-enrichment. However, an extended sample of such sources is needed. Also, without the proof of the $\alpha$ enhancement, we cannot be sure of the SN accretion as the reason of the r-enrichment. Surely, another explanation can be the galactic distribution or local enrichment of r-process elements in the ISM.
If this is the reason, we expect the abundances follow the solar r-process pattern \citep{2000ApJ...544..302B} like in the case of CS 22892-052 \citep{2003ApJ...591..936S}. The s-process pattern may deviate due to the evolved state of our source. Future observations with higher resolution and signal to noise will show a clearer picture.\\
Concluding, we can neither confirm nor reject whether HIP 13692 witnessed a SN in a former binary system due to its already evolved state.
\begin{table}
\centering
\caption{Elemental abundances of HIP 13692 in solar units. 
The errors are the $1\sigma$ uncertainties from the line to line scattering. The uncertainties for elements with only one measurement is inferred from the typical uncertainties mentioned in \citet{2012MNRAS.423.3268K}. The last column shows the number of lines \# used for the measurements.}\label{tab:abundance}
\begin{tabular}{c o{5.5} c}
\toprule
Ion	& \multicolumn{1}{c}{[x/H]}	& \#	\\\midrule
Li I	& 0{.}89+0{.}20	& 1	\\
C I	& -0{.}06+0{.}07	& 6	\\
N I	& 0{.}34+0{.}03	& 3	\\
O I	& -0{.}15+0{.}20	& 1	\\
Na I	& 0{.}14+0{.}20	& 1	\\
Mg I	& -0{.}18+0{.}15	& 3	\\
Al I	& 0{.}11+0{.}07	& 4	\\
Si I	& 0{.}07+0{.}07	& 13	\\
S I	& 0{.}02+0{.}10	& 3	\\
Ca I	& -0{.}10+0{.}14	& 4	\\
Ti I	& 0{.}10+0{.}13	& 5	\\
Ti II  &  0{.}03+0{.}20 &    1\\
V I	& 0{.}15+0{.}10	& 3	\\
V II	& 0{.}00+0{.}03	& 2	\\
Cr I  &   -0{.}01+0.20 &    3 \\
Cr II &   0{.}08+0.20  &   1 \\
Mn I	& -0{.}02+0{.}05	& 3	\\
Fe I	& 0{.}02+0{.}14	& 55	\\
%Co I   &  -0{.}33+0{.}20  &   1 \\
Fe II	& 0{.}03+0{.}08	& 11	\\
Ni I	& 0{.}01+0{.}13	& 13	\\
Cu I  &   0{.}02+0{.}20 &    1 \\
YI I  &   0{.}35+0{.}20 &    1 \\
Zr II &   0{.}07+0{.}20 &    1 \\
Ce II &   0{.}18+0{.}15 &    2 \\
Pr II &   0{.}02+0{.}12 &    2 \\
Nd II &   0{.}36+0{.}08 &    3 \\
Eu II &   0{.}39+0{.}20 &    1 \\
Gd II &   0{.}42+0{.}20 &    1 \\\bottomrule
\end{tabular}
\end{table}
% 

%______________________________________________________________

\section{Conclusions}\label{sec:concl}

We searched for the origin of the young pulsar \psrjna{}. In order to account for the uncertainties in the observables as well as the unknown radial velocity of NSs, we performed Monte Carlo simulations and evaluated the outcome statistically. 

\begin{figure}
\centering
\includegraphics[width=0.45\textwidth, viewport= 20 210 580 625]{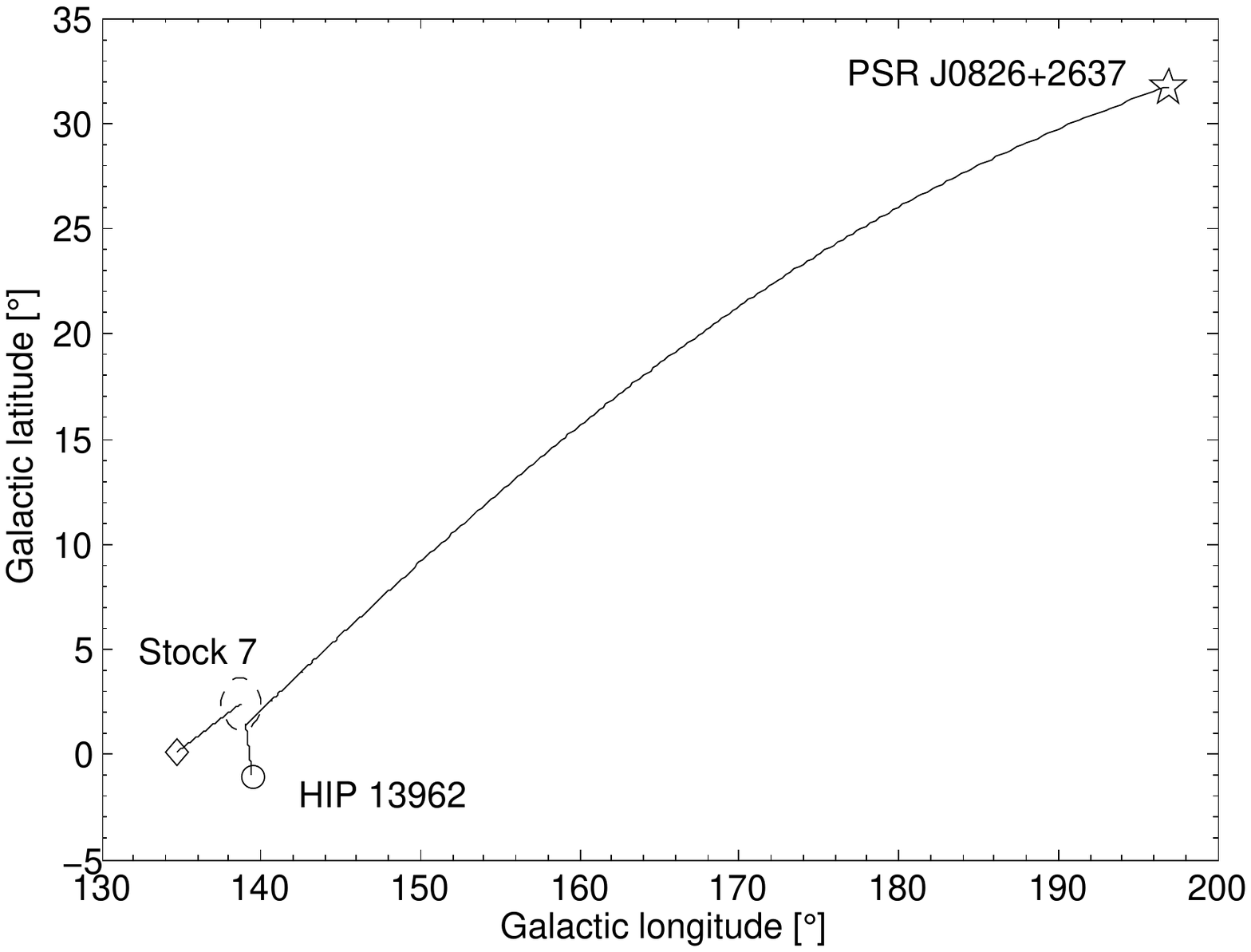}
\caption{Past trajectories for \psrjna{}, the runaway star HIP 13962 and the cluster Stock 7 projected on a Galactic coordinate system (for a particular set of input parameters consistent with \autoref{tab:0826predparmain}) traced back for $\unit[3]{Myr}$. Present positions are marked with a star for the NS, an open circle for the runaway star and a diamond for the cluster. The dashed circle reflects a $\unit[30]{pc}$ radius (see footnote 1).}\label{fig:paths}
\end{figure}

We found that \psrjna{} was possibly ejected from the small cluster Stock 7 with the G0Ia runaway star HIP 13962 being its possible former companion (\autoref{fig:paths}). The predicted kinematic age of the NS is $\unit[3{.}0\pm0{.}6]{Myr}$. This is comparable to the spin-down age and suggests a braking index of $4{.}3^{+0{.}8}_{-0{.}6}$ assuming that the initial spin period was negligible and no glitches occurred.\\
We cannot prove an $\alpha$ enhancement of the runaway star mainly due to the highly evolved state of the star. Even if there had been an enrichment, convective mixing concealed it. However, the binary SN scenario may be supported by the overabundance of r-process elements such as Eu and Gd that were possibly ejected during the SN and accreted by the runaway star. Also, the high rotational velocity of $v\sin i\sim\unit[29]{km/s}$ is consistent with former binary evolution. However, we cannot prove that HIP 13962 gained its runaway status in a past SN event and stress that the star remains a former companion candidate to the NS progenitor.\\
The predicted radial velocity of \psrjna{} is almost zero, implying a 3D velocity of $\unit[183^{+39}_{-32}]{km/s}$ and an inclination to the line-of-sight of the pulsar's motion of $i=\unit[87\pm11]{deg}$. This 3D velocity vector can be used to further investigate the orientation between the pulsar's velocity and spin vectors, if the 3D spin vector was also known. \citet{2004ApJ...615..908B} find an X-ray pulsed fraction of $49\pm22$ per cent ($2\sigma$). In principle, with phase-resolved spectroscopy, it is possible to determine the angle between the rotational and magnetic field axes although the current data is not sufficient.

\section*{Acknowledgments}

We are greatful to \L. Bukowiecki for discussing the age of the cluster Stock 7.

NT acknowledges Carl-Zeiss-Stiftung for a scholarship.
We acknowledge partial support from DFG in the SFB/TR-7 Gravitational
Wave Astronomy.
This work has made use of the Australia Telescope National Facility (ATNF) pulsar database and the Simbad database, operated
at the Centre de Donn\'ees astronomiques de Strasbourg (CDS).
\bibliographystyle{mn2e}
\bibliography{bib_0826Paper}

\label{lastpage}

\end{document}